\documentclass[preprint,showpacs,preprintnumbers,amsmath,amssymb]{revtex4}
\usepackage{graphicx}
\usepackage{dcolumn}
\usepackage{bm}

\newcommand{\be}{\begin{equation}}
\newcommand{\en}{\end{equation}}
\newcommand{\bea}{\begin{eqnarray}}
\newcommand{\ena}{\end{eqnarray}}

\begin{document}


\title{ Warm inflation in the DGP brane-world model }

\author{Sergio del Campo}
 \email{sdelcamp@ucv.cl}
\affiliation{ Instituto de F\'{\i}sica, Pontificia Universidad
Cat\'{o}lica de Valpara\'{\i}so, Casilla 4059, Valpara\'{\i}so,
Chile.}
\author{Ram\'on Herrera}
\email{ramon.herrera@ucv.cl} \affiliation{ Instituto de
F\'{\i}sica, Pontificia Universidad Cat\'{o}lica de
Valpara\'{\i}so, Casilla 4059, Valpara\'{\i}so, Chile.}

\date{\today}

\begin{abstract}
 Warm inflationary universe models on a warped Dvali-Gabadadze-Porrati
  brane are studied. General
 conditions required for these models to be realizable are derived and
 discussed. By using an effective exponential potential we develop  models for
  constant and variable dissipation
 coefficient ratio $r=\frac{\Gamma}{3\,H}$.  We  use recent astronomical
 observations  for constraining the parameters appearing in our models.
\end{abstract}

\maketitle

\section{Introduction}

It is well  know that warm inflation, as opposed to the
conventional cool inflation, presents the attractive feature that
it avoids the reheating period \cite{warm}. In these kind of
models dissipative effects are important during the inflationary
period, so that radiation production occurs concurrently together
with the inflationary expansion. If the radiation field is in a
highly excited state during inflation, and this has a strong
damping effect on the inflaton dynamics, then, it is found a
strong regimen of  warm inflation. Also, the dissipating effect
arises from a friction term which describes the processes of the
scalar field dissipating into a thermal bath via its interaction
with other fields. Warm inflation shows how thermal fluctuations
during inflation may play a dominant role in producing the initial
fluctuations  necessary for Large-Scale Structure (LSS) formation.
In these kind of models the density fluctuations arise from
thermal rather than quantum fluctuations \cite{62526}. These
fluctuations have their origin in the hot radiation and influence
the inflaton through a friction term in the equation of motion of
the inflaton scalar field \cite{1126}. Among the most attractive
features of these models, warm inflation end when the universe
heats up to become radiation domination; at this epoch the
universe stops inflating and "smoothly" enters in a radiation
dominated Big-Bang phase\cite{warm}. The matter components of the
universe are created by the decay of either the remaining
inflationary field or the dominant radiation field
\cite{taylorberera}.

In the Dvali-Gabadadze-Porrati (DGP) model \cite{DGP} the induced
gravity brane-world was put forward as an alternative to the
Randall-Sundrum (RS) one-brane model \cite{RS}, in which general
relativity was recovered, also despite an infinite extra
dimension, but without warping in 5-dimensional Minkowski
space-time. In the DGP model, the gravitational behaviors on the
brane are commanded by the competition between the 5D curvature
scalar in the bulk and the 4D curvature scalar on the brane. In
contrast to the RS case with high energy modifications to general
relativity, the DGP brane produced a low energy modification.  In
the DGP model, according to the embedding of the brane in the
bulk, there appear two branches of background solutions. For a
review of the phenomenology of DGP model, see\cite{Lue} and
inflation models and reheating in this scenario were studied in
\cite{Bou,Pap, Rong, Rong2}.

Usually, in any  models of warm  inflation, the scalar field,
which drives inflation, it is the standard inflaton field. As far
as we know, a model in which warm inflation on a warped DGP brane
has not been yet studied. The main goal of the present work is to
investigate the possible realization of a  warm inflationary
universe model, where the energy densities (inflaton-radiation)
are confined to the brane in DGP model. In this way, we study
warm-DGP model and the cosmological perturbations, which are
expressed in term of different parameters appearing in our model.
These parameters are constrained from the WMAP three year data
\cite{WMAP}.

The outline of the paper is a follows. The next section presents a
short review in the Friedmann equation on the warped DGP inflation
model. In Section \ref{secti} we present the warm inflationary
phase on the DGP brane. Section \ref{sectpert} deals with the
scalar and tensor perturbations, respectively.  In Section
\ref{exemple} we use an exponential potential for obtaining
explicit expression for our models. Finally, Sec.\ref{conclu}
summarizes our findings. We chose units so that $c=\hbar=1$.

\section{Friedmann equation on the warped DGP brane}

We start by writing down  the Friedmann equation on the warped DGP
brane, by using the  Friedmann-Robertson-Walker (FRW) metric. It
becomes
 \be
 \label{H1}
 H^2+{k\over a^2}={1\over 3\mu^2}\Bigl[\,\rho+\rho_0\bigl(1 +
\epsilon{\cal A}(\rho, a)\bigr)\,\Bigr],
 \en
 where  $H$ is the Hubble parameter, $a$ represents the scale factor,
 $\rho$ is the total energy density, and
 $k$ is the constant curvature of the three-space of the  FRW metric.
 The $\mu$ parameters   denotes the strength of the induced gravity term on the
 brane. Also, the  $\epsilon$ parameter becomes  either $+1$ or $-1$
 representing  the two branches of this model. For $\epsilon=-1$ the brane tension
 can be assume to be positive, while for $\epsilon=+1$  this is negative.
   ${\cal A}$ is
defined by

\be
 {\cal A}=\left[{\cal A}_0^2+{2\eta\over
\rho_0}\left(\rho-\mu^2 {{\cal E}_0\over a^4}
\right)\right]^{1\over 2},
 \en
where
 \be
 {\cal A}_0=\sqrt{1-2\eta{\mu^2\Lambda\over \rho_0}},\,\;\;\rho_0=m_\lambda^4+6{m_5^6\over
 \mu^2},\;\;
\eta={6m_5^6\over \rho_0\mu^2} ~~~(0<\eta\leq 1),\label{eq4}
 \en
 and $\Lambda$ is defined by
 \be
 \label{eq5}
\Lambda={1\over 2} ({}^{(5)}\Lambda+{1\over
6}\kappa_5^4\lambda^2),
 \en
 where $\kappa_5$ is the 5-dimensional Newton constant, ${}^{(5)}\Lambda$ is the 5-dimensional cosmological constant
 in the bulk,  and
 $\lambda$ is the brane tension.
 Note  that  there are three mass scales, $\mu$,
$m_\lambda=\lambda^{1/4}$ and $m_5=\kappa_5^{-2/3}$. Also, the
quantities  ${\cal E}_0$ is a constant related to Weyl radiation.
Since we are interested in the inflationary dynamics of the model,
 we will neglect the curvature term and the dark radiation term. We shall restrict ourselves to
 the RS critical
case, i.e. $\Lambda=0$. Then  Eq.(\ref{H1}) becomes
 \be
 \label{newfried}
 H^2=\frac{1}{3\mu^2}\left[\rho+\rho_0+\epsilon \rho_0 (1
 +\frac{2\eta\rho}{\rho_0})^{1/2}\right].
 \en

 Note that in the ultra high energy limit where
 $ \rho \gg \rho_0 \gg m_{\lambda}^4$,   Eq.(\ref{newfried})
 results to be
  \begin{equation}
 \label{high}
 H^2 = \frac{1}{3\mu^2}\left( \rho +\epsilon \sqrt{2\rho
 \rho_0}\right).
 \end{equation}
 In the
 intermediate energy region where $\rho \ll \rho_0 $ but $\rho \gg
 m_{\lambda}^4$, for the branch with $\epsilon =-1$, the Friedamnn
 equation reads $H^2  = \frac{m_{\lambda}^4}{18m_5^6}\left( \rho
+\frac{\rho^2}{2m_{\lambda}^4} -\frac{\mu^2
m_{\lambda}^4}{6m_5^6}\rho -\frac{\mu^2}{4m_5^6}\rho^2\right).$
  Finally,  when $\rho \ll m_{\lambda}^4 \ll \rho_0$ i.e. in low energy limit,
 Eq.(\ref{newfried}) becomes
 \begin{equation}
 \label{low}
 H^2 = \frac{1}{3\mu^2_p}\left [\rho + {\cal O}\left
(\frac{\rho}{\rho_0}\right )^2\right],
 \end{equation}
 where  $\mu_p$ is the effective 4-dimensional Planck mass and is given by $\mu^2_p= \mu^2/(1-\eta)$.

In the following  we will consider a total energy density
$\rho=\rho_\phi+\rho_\gamma$ where $\phi$ corresponds to a
self-interacting scalar field with  energy density, $\rho_\phi$,
given by, $\rho_\phi=\frac{1}{2}\dot{\phi}^2+V(\phi)$ and
$\rho_\gamma$ represents  the radiation energy density.

\section{Warm-DGP Inflationary phase \label{secti}}

 The dynamics of the
cosmological model in the warm-DGP inflationary scenario is
described by the equations
 \be \ddot{\phi}+\,3H \;
\dot{\phi}+V_{,\,\phi}=-\Gamma\;\;\dot{\phi}, \label{key_01}
 \en
and \be \dot{\rho}_\gamma+4H\rho_\gamma=\Gamma\dot{\phi}^2
.\label{3}\en Here $\Gamma$ is the dissipation coefficient and it
is responsible of the decay of the scalar field into radiation
during the inflationary era. $\Gamma$ can be assumed to be a
constant or a function of the scalar field $\phi$ or the
temperature $T$ or both \cite{warm}. Here, we will take $\Gamma$
to be a function of $\phi$ only. At the end subsection B of
section V, we briefly describe the situation in which $\Gamma =$
const. In the near future we hope to study more realistic models
in which $\Gamma$ not only depends on $\phi$ but also on $T$,
expression which could be derived from first principles via
Quantum Field Theory approach \cite{Moss,Bastero}. On the other
hand, $\Gamma$ must satisfies $\Gamma=f(\phi)>0$ by the Second Law
of Thermodynamics. Dots mean derivatives with respect to time and
$V_{,\,\phi}=\partial V(\phi)/\partial\phi$.

During the inflationary epoch the energy density associated to the
scalar field is of the order of the potential, i.e. $\rho_\phi\sim
V$, and dominates over the energy density associated to the
radiation field, i.e. $\rho_\phi>\rho_\gamma$.  Assuming the set
of slow-roll conditions, i.e. $\dot{\phi}^2 \ll V(\phi)$, and
$\ddot{\phi}\ll (3H+\Gamma)\dot{\phi}$ \cite{warm}, the Friedmann
equation (\ref{newfried})  reduces  to
\begin{eqnarray}
H^2\approx\frac{1}{3\mu^2}\left[V+\rho_0+\epsilon \rho_0 ({\cal
A}_0^2
 +\frac{2\eta\;V}{\rho_0})^{1/2}\right],\label{inf2}
\end{eqnarray}
and  Eq. (\ref{key_01}) becomes
\begin{equation}
3H\left[\,1+r \;\right ] \dot{\phi}\approx-V_{,\,\phi},
\label{inf3}
\end{equation}
where $r$ is the rate defined as
\begin{equation}
 r=\frac{\Gamma}{3H }.\label{rG}
\end{equation}
For the high (weak) dissipation  regimen, we have $r\gg 1$ ($r<
1$).

We also consider that  during  warm inflation the radiation
production is quasi-stable, i.e. $\dot{\rho}_\gamma\ll 4
H\rho_\gamma$ and $ \dot{\rho}_\gamma\ll\Gamma\dot{\phi}^2$.  From
Eq.(\ref{3}) we obtained that the energy density of the radiation
field becomes
 \begin{equation}
\rho_\gamma=\frac{\Gamma\dot{\phi}^2}{4H},\label{rh}
\end{equation}
which  could be written as $\rho_\gamma= \sigma T_r^4$, where
$\sigma$ is the Stefan-Boltzmann constant and $T_r$ is the
temperature of the thermal bath. By using Eqs.(\ref{inf3}),
(\ref{rG}) and (\ref{rh}) we get
\begin{equation}
\rho_\gamma=\sigma\,T_r^4=\frac{r\;\mu^2}{4\;(1+r)^2}\left[\frac{V_{,\,\phi}^2}{\left[V+\rho_0+\epsilon
\rho_0 ({\cal A}_0^2
 +\frac{2\eta\;V}{\rho_0})^{1/2}\right]}\right]
.\label{rh-1}
\end{equation}
Introducing the dimensionless slow-roll parameter we get
\begin{equation}
\varepsilon\equiv-\frac{\dot{H}}{H^2}=\frac{\mu^2}{2}\frac{1}{(1+r)}
\left[\frac{V_{,\,\phi}}{V}\right]^2\;\left[\frac{1+\epsilon\eta({\cal
A}_0^2+2\eta\;V/\rho_0)^{-1/2}}{\left[
1+\frac{\rho_0}{V}(1+\epsilon({\cal
A}_0^2+2\eta\frac{V}{\rho_0})^{1/2})\right]^2}\right],\label{ep}
\end{equation}
and the second slow-roll parameter  $\alpha$ becomes
\begin{equation}
\alpha\equiv-\frac{\ddot{H}}{H
\dot{H}}\simeq\,\frac{V_{\,,\,\phi\,\phi}}{3\,(1+r)\,H^2}.\label{eta}
\end{equation}

We see that for $r=0$ (or $\Gamma=0$), the parameters
$\varepsilon$ and $\alpha$ given by Eqs.(\ref{ep}) and (\ref{eta})
respectively, are reduced  to the typical expression for cool
inflation in the DGP brane\cite{Rong}. Note that the term in the
bracket of Eq.(\ref{ep}) is the correction to the standard warm
inflationary model. Also, this term loses the condition for
inflation, i.e. $\varepsilon\ll 1$, and when $\epsilon=1$ and
tightens the condition when $\epsilon=-1$, in contrast to the RS
model where the correction term always loses the inflationary
condition.

It is possible to find a relation between the energy densities
$\rho_\gamma$ and $\rho_\phi$ given by
$$
\rho_\gamma=\frac{r}{2(1+r)}\varepsilon\left[\frac{\rho_\phi+\rho_0(1+\epsilon[{\cal
A}_0^2+2\eta\frac{\rho_\phi}{\rho_0}]^{1/2})}{1+\epsilon\eta({\cal
A}_0^2+2\eta\;\rho_\phi/\rho_0)^{-1/2}}\right]\simeq\frac{r}{2(1+r)}\varepsilon\left[\frac{V+\rho_0(1+\epsilon[{\cal
A}_0^2+2\eta\frac{V}{\rho_0}]^{1/2})}{1+\epsilon\eta({\cal
A}_0^2+2\eta\;V/\rho_0)^{-1/2}}\right].
$$
Recall that during inflation  the energy density of the scalar
field becomes dominated by the potential energy, i.e.
$\rho_\phi\sim V$.

The condition  which the warm inflation epoch on a  DGP Brane
could take place can be summarized with the parameter
$\varepsilon$ satisfying  the inequality  $\varepsilon<1.$ This
condition is analogue to the requirement that  $\ddot{a}> 0$. The
condition given above is rewritten in terms of the densities by
using $\rho_\gamma$, we get
\begin{equation}
\left[\frac{\rho_\phi+\rho_0(1+\epsilon[{\cal
A}_0^2+2\eta\frac{\rho_\phi}{\rho_0}]^{1/2})}{1+\epsilon\eta({\cal
A}_0^2+2\eta\;\rho_\phi/\rho_0)^{-1/2}}\right]>
\frac{2(1+r)}{r}\;\rho_\gamma.\label{cond}
\end{equation}

Inflation ends when the universe heats up at a time when
$\varepsilon\simeq 1$, which implies
\begin{equation}
\left[\frac{V_{,\,\phi}}{V}\right]^2\;\left[\frac{1+\epsilon\eta({\cal
A}_0^2+2\eta\;V/\rho_0)^{-1/2}}{\left[
1+\frac{\rho_0}{V}(1+\epsilon({\cal
A}_0^2+2\eta\frac{V}{\rho_0})^{1/2})\right]^2}\right]\simeq\frac{2}{\mu^2}(1+r).
\end{equation}
The number of e-folds at the end of inflation is given by
\begin{equation}
N=-3\int_{\phi_{*}}^{\phi_f}\frac{H^2}{V_{,\,\phi}}(1+r)
d\phi'.\label{N}
\end{equation}

In the following, the subscripts  $*$ and $f$ are used to denote
to the epoch when the cosmological scales exit the horizon and the
end of  inflation, respectively.

\section{Perturbations\label{sectpert}}

In this section we will study the scalar and tensor perturbations
for our model. Note that in the  case of scalar perturbations the
scalar and the radiation fields are interacting. Therefore,
isocurvature (or entropy) perturbations are generated besides of
the adiabatic ones. This occurs because warm inflation can be
considered as an inflationary model with two basics fields
\cite{Jora1,Jora}. In this context dissipative effects  can
produce a variety of spectral, ranging between red and blue
\cite{62526,Jora}, and thus producing the running blue to red
spectral suggested by WMAP three-year data\cite{WMAP}.

As argued in Ref.\cite{Rong} for the DGP brane (see also
\cite{Bou,Pap,Nunes}), the density perturbation could be written
as $\delta_H=\frac{2}{5}\frac{H}{\dot{\phi}}\,\delta\phi$
\cite{Liddle}.

From Eqs.(\ref{inf3}) and (\ref{rG}), the latter equation becomes
\begin{equation}
\delta_H^2=\frac{36}{25}\frac{H^4\,r^2}{V_{\,,\,\phi}^2}\,\delta\phi^2.\label{331}
\end{equation}

The  scalar field presents  fluctuations which are due to the
interaction between the scalar and the radiation fields.  In the
case of high dissipation, the dissipation coefficient $\Gamma$ is
much greater that the  rate expansion $H$ , i.e. $r=\Gamma/3H\gg
1$ and following  Taylor and Berera\cite{Bere2}, we can write
\begin{equation}
(\delta\phi)^2\simeq\,\frac{k_F\,T_r\,}{2\,\pi^2},\label{del}
\end{equation}
where  the wave-number $k_F$ is defined by $k_F=\sqrt{\Gamma
H/V}=H\,\sqrt{3 r}\geq H$, and corresponds to the freeze-out scale
at which dissipation damps out to the thermally excited
fluctuations. The freeze-out wave-number $k_F$ is defined at the
point where the inequality $V_{,\,\phi\,\phi}< \Gamma H$, is
satisfied \cite{Bere2}.

From Eqs. (\ref{331}) and (\ref{del}) it follows that
\begin{equation}
\delta^2_H\approx\;\frac{18\,\sqrt{3}}{25\,\pi^2}\,
\left[\frac{H^5\,T_r\,r^{5/2}}{V_{\,,\,\phi}^2}\right].\label{dd}
\end{equation}

The scalar spectral index $n_s$ is given by $ n_s -1 =\frac{d
\ln\,\delta^2_H}{d \ln k}$,  where the interval in wave number is
related to the number of e-folds by the relation $d \ln k(\phi)=-d
N(\phi)$. From Eq.(\ref{dd}), we get
\begin{equation}
n_s  \approx\,
1\,-\,\left[5\widetilde{\varepsilon}-2\widetilde{\alpha}
+\frac{5\,r_{\,,\,\phi}\,V_{\,,\,\phi}}{6\,r^2\,H^2}\right]\approx\,1-\frac{5\,\widetilde{\varepsilon}}{2}+2\widetilde{\alpha}
-\frac{5\;V_{\,,\,\phi}\,\Gamma_{\,,\phi}}{2\,H\,\Gamma^2},\label{ns1}
\end{equation}
where, the  slow-roll parameters $\widetilde{\varepsilon}$ and
$\widetilde{\alpha}$, (for $r\gg 1$) are given by
\begin{equation}
\widetilde{\varepsilon}\approx\frac{\mu^2}{2\,r}\,
\left[\frac{V_{,\,\phi}}{V}\right]^2\;\left[\frac{1+\epsilon\eta({\cal
A}_0^2+2\eta\;V/\rho_0)^{-1/2}}{\left[
1+\frac{\rho_0}{V}(1+\epsilon({\cal
A}_0^2+2\eta\frac{V}{\rho_0})^{1/2})\right]^2}\right],\,\;\mbox{and}\;\;
\,\,\widetilde{\alpha}\approx\,\frac{V_{\,,\,\phi\,\phi}}{3\,r\,H^2},
\end{equation}
respectively.

One of the interesting features of the three-year data set from
WMAP is that it hints at a significant running in the scalar
spectral index $dn_s/d\ln k=\alpha_s$ \cite{WMAP}. Dissipative
effects themselves can produce a rich variety of spectra ranging
between red and blue \cite{62526,Jora}. From Eq.(\ref{ns1}) we
obtain that the running of the scalar spectral index becomes

\begin{equation}
\alpha_s=\frac{15\,r\,V_{,\,\phi}}{\Gamma^2}\;\left[\widetilde{\varepsilon}_{,\,\phi}-
\frac{2\,\widetilde{\alpha}_{,\,\phi}}{5}-\frac{(\ln
r)_{,\,\phi}}{2}\,[\widetilde{\alpha}-2\widetilde{\varepsilon}]-
\frac{3\,V_{,\,\phi}}{2\,\Gamma^2}\,\left(r_{,\phi\phi}-2\frac{r_{,\,\phi}^2}{r}\right)\right].\label{dnsdk}
\end{equation}
In models with only scalar fluctuations the marginalized value for
the derivative of the spectral index is approximately $-0.05$ from
WMAP-three year data only \cite{WMAP}.

As it was mentioned in Ref.\cite{Bha} the generation of tensor
perturbations during inflation would  produce stimulated emission
in the thermal background of gravitational wave. This process
changes the power spectrum of the tensor modes by an extra
temperature dependently  factor given by $\coth(k/2T)$. The
corresponding spectrum  becomes
\begin{equation}
A^2_g=\frac{16\pi}{\mu^2}\left(\frac{H}{2\pi}\right)^2\,\coth\left[\frac{k}{2T}\right]
\simeq\frac{4}{3\pi\mu^4}\,\left[V+\rho_0+\epsilon \rho_0 ({\cal
A}_0^2
 +\frac{2\eta\;V}{\rho_0})^{1/2}\right]\coth\left[\frac{k}{2T}\right],\label{ag}
\end{equation}
where the spectral index $n_g$, results to be given by $
n_g=\frac{d}{d\,\ln k}\,\ln\left[
\frac{A^2_g}{\coth[k/2T]}\right]=-2\,\varepsilon$.  Here, we have
used that  $A^2_g\propto\,k^{n_g}\,\coth[k/2T]$ \cite{Bha}.

For $r\gg1$ and from expressions (\ref{dd}) and (\ref{ag}) we may
write  the tensor-scalar ratio as
\begin{equation}
R(k)=\left.\left(\frac{A^2_g}{P_{\cal R}})\right)\right|_{\,k_*}
\simeq\left.\frac{8\pi}{9\sqrt{3}\mu^2}
\left[\frac{V_{\,,\,\phi}^2}{T_r\,H^3\,r^{5/2}}\,\coth\left(\frac{k}{2T}\right)\right]\right|_{\,k=k_*}.
\label{Rk}\end{equation} Here, $\delta_H\equiv\,2\,P_{\cal
R}^{1/2}/5$ and  $k_*$  is referred to $k=Ha$, the value when the
universe scale  crosses the Hubble horizon  during inflation.

Combining  WMAP three-year data\cite{WMAP} with the Sloan Digital
Sky Survey  (SDSS) large scale structure surveys \cite{Teg}, it is
found an upper bound for $R$ given by $R(k_*\simeq$ 0.002
Mpc$^{-1}$)$ <0.28\, (95\% CL)$, where $k_*\simeq$0.002 Mpc$^{-1}$
corresponds to $l=\tau_0 k\simeq 30$,  with the distance to the
decoupling surface $\tau_0$= 14,400 Mpc. The SDSS  measures galaxy
distributions at red-shifts $a\sim 0.1$ and probes $k$ in the
range 0.016 $h$ Mpc$^{-1}$$<k<$0.011 $h$ Mpc$^{-1}$. The recent
WMAP three-year results give the values for the scalar curvature
spectrum $P_{\cal R}(k_*)\equiv\,25\delta_H^2(k_*)/4\simeq
2.3\times\,10^{-9}$ and the scalar-tensor ratio $R(k_*)=0.095$. We
will make use of these values  to set constrains on the parameters
for our model.

\section{Exponential potential in the high dissipation approach \label{exemple}}
Let us consider  an inflaton scalar field $\phi$ on the brane with
exponential potential. This potential occur naturally in some
fundamental theories such as string/M theories and it is
intensively studied in   DGP models.

We write for the exponential potential as $V=V_0
e^{-\sqrt{2/p}\;\frac{\phi}{\mu}}$, where $V_0$ and $p$  are two
constants, and $V(\phi)\longrightarrow$ 0 as $\phi\longrightarrow
\infty$, which means $p> 0$.  An estimation of these parameters
are give for cool inflation on a DGP brane in Ref.\cite{Rong}. In
the following, we will restrict ourselves to the high dissipation
regimen i.e. $r\gg 1$.

\subsection{ $\Gamma \propto \left[v(\phi)+
1+\epsilon({\cal A}_0^2+2\eta\,v(\phi))^{1/2})\right]^{1/2}$
case.\label{GammaConst}}

This choice is motivated by the fact that $r$ becomes a constant.
By using the exponential potential, we find that the slow-roll
parameters become
$$
\widetilde{\varepsilon}=\frac{1}{p\,r_0}\,
\;\left[\frac{1+\epsilon\eta({\cal A}_0^2+2\eta\;v)^{-1/2}}{\left[
1+\frac{1}{v}(1+\epsilon({\cal
A}_0^2+2\eta\,v)^{1/2})\right]^2}\right],\label{eta1} \mbox{and}
\, \widetilde{\alpha}=\frac{2}{p\,r_0}\, \;\left[\frac{v}{\left[v+
1+\epsilon({\cal A}_0^2+2\eta\,v)^{1/2})\right]^2}\right],$$
 where
we have define $v=\frac{V}{\rho_0}$.

By integrating Eq.(\ref{N}) the number of e-folds results in $
N=\frac{p\,r_0}{2}\;[h(v_f)-h(v_*)]$,  where
\begin{equation}
h(v)=-\left[\ln v-v^{-1}+\left(Wv^{-1}+\frac{2\eta}{{\cal
A}_0}\tanh^{-1}(W/{\cal A}_0)\right)\right],
\end{equation}
and $W=W(v)=({\cal A}_0^2+2\eta\,v)^{1/2}$.

In section \ref{secti} we specified that the inflationary phase
can exit naturally without any other mechanism in the branch
$\epsilon=-1$. In this case   the condition for  inflation end,
$\varepsilon=1$,  gives a quintuple equation  that can not be
solved  analytically. Note that the same happens in cool inflation
for the DGP brane \cite{Rong}.

From Eq.(\ref{dd}), when $r\gg 1$, we obtain that the scalar power
spectrum becomes
\begin{equation}
P_{\cal R}(k)\approx\left.
\;\frac{p\;r_0^{5/2}}{4\,\pi^2}\,\left[\frac{\rho_0^{1/2}\,T_r\;v^{1/2}}{\mu^3}\right]
\;\left[1+\frac{1}{v}\left(1+\epsilon\;W\right)\right]^{5/2}\right|_{\,k=k_*},\label{ppp}
\end{equation}
and from Eq.(\ref{Rk}) the tensor-scalar ratio, is given by
\begin{equation}
R(k)\approx\;\left.\frac{16\,\pi}{3}\left[\frac{\rho_0^{1/2}\;v^{1/2}}{\mu\;p\;r_0^{5/2}\,T_r}
\right]\;\;\left[1+\frac{1}{v}\left(1+\epsilon\;W\right)\right]^{-3/2}
\;\coth\left[\frac{k}{2T}\right]\right|_{\,k=k_*}.\label{rrrr}
\end{equation}

By using the WMAP three year data where $P_{\cal R}(k_*)\simeq
2.3\times 10^{-9}$,  $R(k_*)=0.095$, and choosing the parameter
 $T\simeq\,T_r$, we obtained from Eqs.(\ref{ppp}) and
(\ref{rrrr}) that $ v_*=2\,\left[\frac{C-1+\eta}{(C-1)^2}\right]$,
where the  $C$ is given by $
C\simeq\frac{5.1\times\,10^{-10}}{v_\mu\;\coth\left[\frac{k_*}{2T_r}\right]},
$ and $v_\mu=V_*/\mu^4$. Recall  that
$\Lambda=0\Longrightarrow\,{\cal{A}}_0=1$ and $\epsilon=-1$.

Now we consider the special case in which we fixe $N=60$, $p=5$
and $r_0=10$. In this special case we obtained that $v_f=0.05$,
$v_*=36$ and $v_\mu=7.5\times 10^{-13}$. Also, we choose the
parameters $k_*=0.002$ Mpc$^{-1}$ and $T\simeq\,T_r\simeq 0.24
\times 10^{16}$ GeV. We also take  $\eta=0.99$, although the model
can be used for any value of $\eta$.

From Eqs.(\ref{eq4}) and (\ref{eq5}) we obtained that the
 tension of the brane, $\lambda$ becomes
$
\frac{\lambda}{\mu^4}=\frac{\rho_0}{\mu^4}\;(1-\eta)\simeq\;2.1\times\,10^{-16},
$
and the 5-dimensional gravitational constant results to be
$
\frac{m_5}{\mu}=\left[\frac{\eta\;\rho_0}{6\,\mu^4}\right]^{1/6}\simeq\;0.0039.
$
Analogously, we get that the cosmological constant $^{5}\Lambda$
in the bulk can be evaluate and it becomes
 $
^{5}\Lambda/\mu^2\simeq 2.1\times 10^{-18}$. We note that the
$\Lambda$, $m_5$ and $^5\Lambda$ parameters become decrease by
two, one and fourth orders of magnitude, respectively, when are
compared with those analogous cool article   related to  inflation
in the DGP brane.




\subsection{ $\Gamma \propto \left[v(\phi)+
1+\epsilon({\cal A}_0^2+2\eta\,v(\phi))^{1/2})\right]^{3/2}$
case.\label{Gammavariable}}

This choice allows us to get $r$ as a function of $H$ in analogy
with that one used by authors of Refs.\cite{yo,yo1}. Therefore, we
take $r$ to be of the form, $r=(3\;\mu^2/\rho_0)H^{2}$, and thus
we find for the slow-roll parameters
$$
\widetilde{\varepsilon}=\frac{1}{p\,v}\,
\;\left[\frac{1+\epsilon\eta({\cal A}_0^2+2\eta\;v)^{-1/2}}{\left[
1+\frac{1}{v}(1+\epsilon({\cal
A}_0^2+2\eta\,v)^{1/2})\right]^3}\right],\,\mbox{and}\;\;
\widetilde{\alpha}=\frac{2}{p\,v}\, \;\left[\frac{1}{\left[v+
1+\epsilon({\cal A}_0^2+2\eta\,v)^{1/2})\right]^2}\right].
$$

The  number of e-folds is given by $
N=\frac{p}{2}\,\left[\Im(v_f)-\Im(v_*)\right]$,  where the
function $\Im(v)$ is defined by
$$
\Im(v)=-\left[v-\left(\frac{1+{\cal
A}_0^2\;\epsilon^2}{v}\right)+2\,(1+\eta\,\epsilon) \ln
v+\left(\epsilon\,W\,\left[4-\frac{2}{v}\right]+\frac{4\,\epsilon({\cal
A}_0^2+\eta)}{{\cal A}_0^2}\tanh^{-1}(W/{\cal A}_0)\right)\right].
$$

The $N$ parameter has to have an appropriated values (60 or so)
 in order to solve the standard cosmological puzzles.
 In order to do so we need the following inequality
  $\Im(v_*)< \Im(v_f)-120/p$ with $\Im(v_f)>120/p$ to be satisfied.
Again,  the condition  for  inflation becomes to an end is
$\varepsilon(v_f)=1$, has roots  that can not be written down in
closed algebraic form.

At the epoch when the cosmological scale exit the horizon, the
dissipation parameter becomes $
r(\phi_*)=r_*\gg\,1\Longrightarrow\;\rho_0\ll\,3\,\mu^2\,H_*^2$,
resulting in the requirement that $
[v_*+1+\epsilon({\cal{A}}_0^2+2\eta\,v_*)^{1/2}]\,\gg\;1 $.

From Eq.(\ref{dd}), and for $r\gg 1$, we obtain that the scalar
power spectrum becomes
\begin{equation}
P_{\cal R}(k)\approx\left.
\;\frac{p}{4\,\pi^2}\,\left[\frac{\rho_0^{1/2}\,T_r\,v^3}{\mu^3}\right]
\;\left[1+\frac{1}{v}\left(1+\epsilon\;W\right)\right]^5\right|_{\,k=k_*},\label{pp}
\end{equation}
 and from Eq.(\ref{Rk}) the
tensor-scalar ratio, is given by
\begin{equation}
R(k)\approx\;\left.\frac{24\,\pi}{p}\left[\frac{\rho_0^{1/2}}{\mu\,T_r\;v^2}
\right]\;\;\left[1+\frac{1}{v}\left(1+\epsilon\;W\right)\right]^{-4}
\;\coth\left[\frac{k}{2T}\right]\right|_{\,k=k_*}.\label{rr}
\end{equation}

By using the WMAP three year data, where $P_{\cal R}(k_*)\simeq
2.3\times 10^{-9}$,  $R(k_*)=0.095$, and choosing the parameter
 $T\simeq\,T_r$, we obtained from Eqs.(\ref{pp}) and
(\ref{rr}) that $ v_*=2\,\left[\frac{C-1+\eta}{(C-1)^2}\right]$,
where now   $C$ is given by $
C\simeq\frac{1.1\times\,10^{-10}}{v_\mu\;\coth\left[\frac{k_*}{2T_r}\right]},
$ and $v_\mu=V_*/\mu^4$. Here, once again we have taken
$\Lambda=0\Longrightarrow\,{\cal{A}}_0=1$ and $\epsilon=-1$.

Let us consider  some numerical situations.  For $N=60$ and
$p=60$, we obtain that $v_f=0.63$, $v_*=8$ and $v_\mu=2.1\times
10^{-13}$, where again we have taken  the parameters $k_*=0.002$
Mpc$^{-1}$, $T\simeq\,T_r\simeq 0.24 \times 10^{16}$ GeV and
$\eta=0.99$.

From Eqs.(\ref{eq4}) and (\ref{eq5}), the tension of the brane
becomes $
\frac{\lambda}{\mu^4}=\frac{\rho_0}{\mu^4}\;(1-\eta)\simeq\;2.6\times\,10^{-16},
$ and the 5-dimensional gravitational constant can be written as
$
\frac{m_5}{\mu}=\left[\frac{\eta\;\rho_0}{6\,\mu^4}\right]^{1/6}\simeq\;0.0041.
$
Analogously, we obtain that $ ^{5}\Lambda/\mu^2\simeq 2.7\times
10^{-18}$. Note that theses values are similar to their analogous
for the case in which $r=r_0$.

\begin{figure}[th]
\includegraphics[width=4.0in,angle=0,clip=true]{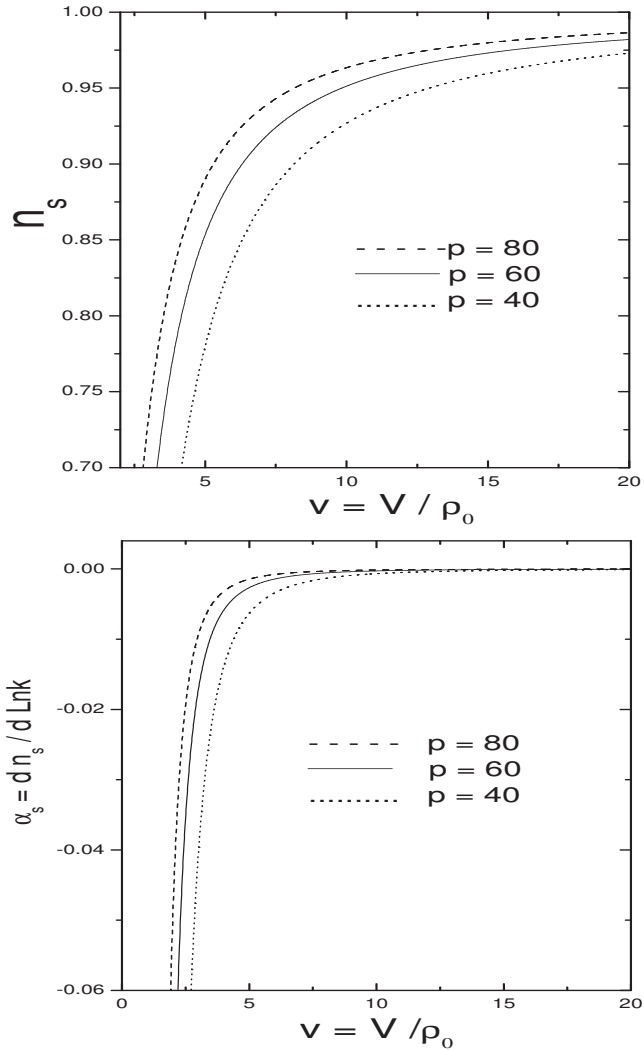}
\vspace{-0.6cm}\caption{ The upper panel show the  evolution of
the scalar spectrum index $n_s$ versus $v=V/\rho_0$, and   the
lower panel show the evolution of the running scalar spectrum
index, $\alpha_s$, versus $v=V/\rho_0$. In both panels we have
used different  values for the parameter $p$, for $r\propto\,H^2$.
Here, we have taken the values $\eta=0.99$, ${\cal{A}}_0=1$, and
$\epsilon=-1$. \label{ns}}
\end{figure}


In Fig.(\ref{ns})  we plot the scalar spectrum index $n_s$  and
the running spectral index $\alpha_s$ versus the ratio
$v=V/\rho_0$.  In doing  this, we have taken three different
values of the parameter p. Note that in considering the $n_s$
versus $v= V/\rho_0$ plot the WMAP-three data favors hight values
of this parameter. However, from the plot $\alpha_s$ versus $v=
V/\rho_0$  we could not say the same, since it seems that any
value of p would give the same value of $\alpha_s$ for hight
enough  $v= V/\rho_0$.

Another interesting case to study is to consider
$\Gamma=\Gamma_0=constant$ i.e. $r\propto H^{-1}$. Analogously as
in the other cases, we take the special case with fixed $N=60$ and
$p=80$. For these values, we have $v_f=0.05$, $v_*=50$,
$v_\mu=1.5\times 10^{-13}$ and $\Gamma_0=8\times 10^{15}$ GeV.
Here, again we have taken  the parameters $k_*=0.002$ Mpc$^{-1}$,
$T\simeq\,T_r\simeq 0.24 \times 10^{16}$ GeV and $\eta=0.99$. From
Eqs.(\ref{eq4}) and (\ref{eq5}), the tension  result to be $
\frac{\lambda}{\mu^4}\simeq\;2.9\times\,10^{-17}, $ and the
5-dimensional gravitational constant becomes $
\frac{m_5}{\mu}\simeq\;0.0028$. Finally, we obtain that $
^{5}\Lambda/\mu^2\simeq 3.0\times 10^{-19}$. Note that these
values are similar to those values in which $\Gamma$ is variable.

\section{Conclusions \label{conclu}}

In this paper we have investigated the  warm inflationary scenario
on a warped DGP brane. In the slow-roll approximation we have
found a general relationship between the radiation and scalar
field energy densities. This has led us to a general criterium for
 warm inflation in DGP brane to occur (see Eq.(\ref{cond})).

Our  specific models are described by  an exponential potential
and we have consider  different  cases for   the dissipation
coefficient, $\Gamma$. In the first case, we took $\Gamma \propto
\left[v(\phi)+ 1+\epsilon({\cal
A}_0^2+2\eta\,v(\phi))^{1/2})\right]^{1/2}$. Here, we have
 found that the condition for inflation to  end presents the
same characteristic that  occurs in cool inflation for the DGP
brane \cite{Rong}, except that it depends on the extra parameter
$r_0$. For the case in which the dissipation coefficient $\Gamma$
is taken to be a function of the scalar field, i.e. $\Gamma
\propto \left[v(\phi)+ 1+\epsilon({\cal
A}_0^2+2\eta\,v(\phi))^{1/2})\right]^{3/2}$, it was possible to
describe an  appropriate   warm inflationary universe model on DGP
brane. In these cases, we have obtained the  explicit expressions
for the corresponding scalar spectrum index and the running of the
scalar spectrum index. We have also study the situation in which
$\Gamma = \Gamma_0= Const.$ Here, we have found that the values
are similar to those found when $\Gamma$ is a function of the
scalar field.

By using the WMAP three year data and consider a special case with
fixed $N$, $p$ and $r_0$,  we have found the values of the
parameters $\lambda$, $^{5}\Lambda$ and $m_5$. Using the above
parameters we will check with a numerical example that the
evolution of the universe really undergoes a 4-dimensional stage,
then a  5-dimensional stage, and finally a 4-dimensional stage
again. From Eqs.(\ref{high}) to (\ref{low}) the  dimensional
transition occurs if $\lambda\mu^2/6 m_5^6\ll 1$ \cite{Lue}. In
our models that we have worked out we have obtained that
$\lambda/\mu^4\simeq 2\times 10^{-16}$, $ m_5/\mu\simeq 0.004$ and
therefore $\lambda\mu^2/6 m_5^6=0.008\ll 1$. Also,
$v_*\sim\rho_*/\rho_0\sim 10$ for these models and thus the
universe inflates in a 4-dimensional stage when the cosmic scale
crossed the Hubble horizon during inflation, since
$\rho_*\gg\rho_0\gg m_\lambda^4$. In the intermediate energy
region (5-dimensional), it  is necessary that
$m_\lambda^4\ll\rho\ll\rho_0$ and using the final value of
$v_f\sim\rho_f/\rho_0$, we find that $\rho_f\ll 1$ and
$\rho_f/m_\lambda^4\sim 10$. Therefore, the inflationary phase
exists in a 5-dimensional stage. Finally, the universe stops
inflating and enters in a radiation dominated Big-Bang phase,  in
which, the energy density decreases very fast and the universe
becomes  4-dimensional again.

We should  note that other properties of this model deserves
further study. For example, we did not study  bispectrum of
density perturbations. While cool inflation typically predicts a
nearly vanishing bispectrum, and hence a small (just a few per
cent) deviation from Gaussianity in density fluctuations -see e.g.
\cite{vanishing}, warm inflation clearly predicts a non-vanishing
bispectrum. We left  this question for a subsequent study of the
 warm-DGP brane-world. Also, a more accurate calculation for the
 density perturbation would be necessary in order to check the validity
of expression  (\ref{331}). We intend
 to return to this point in the near future by working  an approach
 analogous to that followed  in Refs.\cite{yo} and \cite{yo1}.

\begin{acknowledgments}
 S.d.C. was
supported by COMISION NACIONAL DE CIENCIAS Y TECNOLOGIA through
FONDECYT grant N$^0$ 1070306. Also, from UCV-DGIP N$^0$
123.787/2007. R.H. was supported by the ``Programa Bicentenario de
Ciencia y Tecnolog\'{\i}a" through the Grant ``Inserci\'on de
Investigadores Postdoctorales en la Academia" \mbox {N$^0$
PSD/06}.
\end{acknowledgments}


\end{document}